\documentclass{IEEEtran}
\usepackage{amsmath, amssymb}
\usepackage{graphicx,cite,mathrsfs}

\newtheorem{construction}{Construction}
\newtheorem{theorem}{Theorem}
\newtheorem{lemma}[theorem]{Lemma}
\newtheorem{corollary}[theorem]{Corollary}

\newcommand{\Legendre}[2]{\left( \frac{#1}{#2} \right)}

\newcommand{\set}[1]{\mathcal{#1}}
\newcommand{\CAC}{\mathsf{CAC}}
\newcommand{\C}{\mathscr{C}}

\title{A General Upper Bound on the Size of Constant-Weight Conflict-Avoiding Codes}

\author{Kenneth W. Shum$^\dagger$,~\IEEEmembership{Member, IEEE}, Wing Shing Wong$^\dagger$,~\IEEEmembership{Fellow, IEEE}, and Chung Shue Chen$^\ddagger,~\IEEEmembership{Member, IEEE.}$  \thanks{This work was supported by a grant from the Research Grants Council of the Hong Kong Special Administrative Region under Project 417909.}
\thanks{$^\dagger$Dept. of Information Engineering,
The Chinese University of Hong Kong,
Shatin, Hong Kong. Email:
wkshum@inc.cuhk.edu.hk, wswong@ie.cuhk.edu.hk.}
\thanks{$^\ddagger$Research Group on Network Theory and Communications (TREC), INRIA-ENS, 75214 Paris, France. Email: Chung-Shue.Chen@inria.fr}}

\begin{document}

\maketitle

\begin{abstract}
Conflict-avoiding codes are used in the multiple-access collision channel without feedback. The number of codewords in a conflict-avoiding code is the number of potential users that can be supported in the system. In this paper, a new upper bound on the size of constant-weight conflict-avoiding codes is proved. This upper bound is general in the sense that it is applicable to all code lengths and all Hamming weights. Several existing constructions for conflict-avoiding codes, which are known to be optimal for Hamming weights equal to four and five, are shown to be optimal for all Hamming weights in general.
\end{abstract}


\section{Introduction}

A set of $k$ binary sequences of length $L$ is called {\em user-irrepressible}~\cite{SWSC} if after cyclically shifting each of them  and stacking them together in a $k\times L$ matrix,  we can always find a $k\times k$ submatrix which is a permutation matrix, regardless of how we shift the $k$ sequences. (Recall that a permutation matrix is a zero-one square matrix with exactly one 1 in each row and each column~\cite[p.25]{HornJohnson}.) A set of $N$ binary sequences is called {\em $(N,k)$-conflict-avoiding}~\cite{Tsybakov02} if every subset of $k$ sequences out of these $N$ sequences is user-irrepressible.

User-irrepressible and conflict-avoiding sequences find applications in
collision channel without feedback~\cite{Tsybakov83, Massey85}. In a system with $k$ active users, the collision channel is a deterministic channel with $k$ inputs and one output defined as follows. Time is assumed to be partitioned into fixed-length time intervals, called {\em slots}. Here, we consider the {\em slot-synchronous} case. In each slot, each user either remains silent or transmits a packet.
If exactly one user transmits in a time slot, then the packet is successfully received and the channel output is the same as the packet sent by that user. If two or more users transmit in the same time slot, a collision occurs and the channel output is an erasure symbol ``*''. If none of the users transmits in a time slot, the time slot is idle.

Suppose that there are $N$ potential users, but at most $k$ of them are active at the same time. This model is applicable to communication system in which traffic is bursty and the users transmit signal intermittently.
We assign statically each of the $N$ users a binary sequence from a set of $(N,k)$-conflict-avoiding sequences. Each active user reads out the assigned sequence periodically, and sends a packet if and only if the value of the sequence is equal to~1. Since there is no timing information other than that for slot synchronization, the starting time of the sequences may  be different and relative delay offsets are incurred. The user-irrepressible property translates to the following non-blocking property:
for each active user we can find  at least one time slot in a period in which this user transmits a packet while the remaining $k-1$ active users are silent, i.e., each active user can transmit at least one packet without collision in $L$ time slots. This provides a worst-case guarantee of bounded delay.

There are two different but complementary design goals in the literature of user-irrepressible and conflict-avoiding sequences. In the first one, we consider the scenario in which all the $N$ users are active, i.e.~$N=k$, and we aim at minimizing the length of the binary sequences while keeping the user-irrepressible property~\cite{UIS08, SWSC}. We can add an inner code, such as  Reed-Solomon code, in order to recover collided packets and enhance system throughput. In the second one, we consider a fixed sequence length and a given number of $k$ active users, and aim at maximizing the total number of potential users that can be supported. Each active may repeatedly sending the same packet in one sequence period. The packet is guaranteed to be received successfully within the duration of a period. This viewpoint is adopted in~\cite{LT05, Levenshtein07, Momihara07b, JMJTT, Mishima09, Momihara07}. In this paper, we consider the second design goal and maximize the number of potential users for a given sequence length. Other coding constructions for multiple access in collision channel without feedback, such as constant-weight cyclically permutable codes, can be found in~\cite{Nguyen92,GyorfiVajda93,MZKZ95}.

The number of ones in a binary sequence is called the {\em  Hamming weight}. It is easy to see that in order to support user-irrepressibility, each active user has to send at least $k$ packets in a period of $L$ time slots, i.e., the Hamming weight of the sequence is at least $k$. Otherwise, if a user sends only $k-1$ packets in a period, we can always arrange the delay offsets of the other $k-1$ users so that all these $k-1$ packets are in collision, violating the property of user-irrepressibility.
In this paper, we focus on the extreme case where all sequences have the same Hamming weight $w$ which equals the number of active users, i.e., $w=k$. This is the minimum weight requirement for user-irrepressibility.
Under the assumption of $w=k$, many works are devoted to determine the maximal number of potential users for Hamming weight equal to three, see e.g.\ \cite{LT05, Levenshtein07, JMJTT, Momihara07b, Mishima09}.  Some optimal constructions for Hamming weight equal to four and five are presented in~\cite{Momihara07}.  However, the maximal number of potential users for general Hamming weight $w$ larger than five is unknown. We address this open question in this paper and provide a general upper bound on the number of potential users for all Hamming weights. An asymptotic version of this general upper bound can be found in~\cite{ShumWong09}.

This paper is organized as follows. We define conflict-avoiding codes and set up some notations in Section~\ref{sec:def}. Three known constructions are described in Section~\ref{sec:known_construction}.  The main  result in this paper is contained in Section~\ref{sec:Kneser}, which provides an upper bound on the number of potential users that can be supported, given the length $L$ and Hamming weight~$w$. In Section~\ref{sec:optimal}, we apply this upper bound to the constructions described in Section~\ref{sec:known_construction}. Optimal CAC with Hamming weight $w \geq 6$ are also given in Section~\ref{sec:optimal}.

\section{Definitions  and Notations} \label{sec:def}
We represent a binary sequence by specifying the time indices where the sequence value is equal to one.
Let $\mathbb{Z}_L = \{0,1,\ldots,L-1\}$ be the set of integers reduced modulo~$L$.
A subset $\set{I}$ of $\mathbb{Z}_L$ is associated with a binary sequence $s(t)$ of length $L$ with Hamming weight $|\set{I}|$, by setting $s(t) = 1$ if and only if $t\in\set{I}$, where $|\set{I}|$ denotes the cardinality of $\set{I}$.
Subsets of $\mathbb{Z}_L$ with cardinality $w$ are called {\em codewords}.

For a codeword $\set{I}$, let
\[
 d(\set{I}) := \{a-b \bmod L:\, a,b\in \set{I}\}
\]
denote the {\em set of differences} between pairs of elements in~$\set{I}$. Since $a$ may equal to $b$ in the definition of $d(\set{I})$, it is obvious that 0 is always an element in $d(\set{I})$.
Let $d^*(\set{I})$ be the set of non-zero differences in $d(\set{I})$,
\[
 d^*(\set{I}) := d(\set{I}) \setminus \{0\}.
\]
It is the set of differences between pairs of {\em distinct} elements in~$\set{I}$.
A collection of $M$ codewords
\[\C = \{\set{I}_1,\set{I}_2,\ldots, \set{I}_M\}
 \]
is called a {\em conflict-avoiding code} (CAC) of length $L$ and weight $w$ if
\[
d^*(\set{I}_j) \cap d^*(\set{I}_k) = \emptyset,
\]
for all $j\neq k$. We use the notation $(L,w)$-$\CAC$ for a conflict-avoiding code of length $L$ and weight $w$. It is easy to see that an $(L,w)$-$\CAC$ with $N$ codewords is equivalent to a set of $(N,w)$-conflict-avoiding sequences mentioned in the introduction. We sometime say that $\set{I}$ is a codeword of weight~$w$.
Since adding a constant to all elements in a codeword $\set{I}$ does not affect the set of differences $d(\set{I})$, we assume without loss of generality that every codeword in a CAC contains the zero element~0 in~$\mathbb{Z}_L$.

Given positive integers $L$ and $w$, consider the class of all CACs with length $L$ and weight $w$. A CAC in this class with maximal number of codewords is called {\em optimal}, and the maximal number of codewords is denoted by $M(L,w)$. The objective of this paper is to derive an upper bound on $M(L,w)$ for all $L$ and~$w$.

{\em Example 1:} $L=15$, $w=3$. The four codewords $\{0, 5, 10\}$,  $\{0,1, 2\}$, $\{0,7,11\}$ and $\{0,6,12\}$ constitute a $(15,3)$-$\CAC$. We can verify that the sets of non-zero differences
\begin{align*}
d^*(\{0,5,10\}) &= \{5,10\} \\
d^*(\{0,1,2\}) &=  \{1,2,13,14\}\\
d^*(\{0,7,11\}) &=  \{4,7,8,11\}\\
d^*(\{0,6,12\}) &= \{3,6,9,12\}
\end{align*}
are disjoint.

{\em Example 2:} $L=26$, $w=4$. Consider the four codewords $\{0,1,2,3\}$, $\{0,4,8,12\}$, $\{0,5,10,15\}$ and $\{0,6,13,19\}$. Since
\begin{align*}
d^*(\{0,1,2,3\}) &= \{1,2,3,23,24,25\} \\
d^*(\{0,4,8,12\}) &= \{4,8,12,14,18,22\} \\
d^*(\{0,5,10,15\}) &= \{5,10,11,15,16,21\} \\
d^*(\{0,6,13,19\}) &= \{6,7,13,19,20\}
\end{align*}
are  disjoint, we have a $(26,4)$-$\CAC$ with four codewords.

{\bf Remark:} From the definition of CAC, it follows directly that for $j\neq k$, the Hamming cross-correlation of the two binary sequences associated with two distinct codewords in an $(L,w)$-$\CAC$ is no more than 1 for any cyclic shift. An $(L,w)$-$\CAC$ can thus be viewed as an $(L,w,1)$-optical orthogonal code (OOC) without any auto-correlation requirement. We refer the readers to, e.g.~\cite{fujihara00}, and the references therein for further information on OOC.

A codeword $\set{I}$ is called {\em equi-difference} if the elements in $\set{I}$ form an arithmetic progression in $\mathbb{Z}_L$, i.e.,
\[
 \set{I} = \{0, g, 2g, \ldots, (w-1)g\}
\]
for some $g\in\mathbb{Z}_L$. In the above equation, the product $jg$ is reduced mod $L$, for $j=2,3,\ldots, (w-1)$. The element $g$ is called a {\em generator} of this codeword. For an equi-difference codeword $\set{I}$ generated by $g$, the  set of differences is equal to
\[
 d(\set{I}) = \{0, \pm g, \pm 2g, \ldots, \pm (w-1)g\}.
\]
We remark that the elements $\pm g$, $\pm 2g, \ldots, \pm (w-1)g$ may not be distinct mod $L$. Hence in general we have $|d^*(\set{I})| \leq 2w-2$, with equality holds if $\pm g$, $\pm 2g, \ldots, \pm (w-1)g$ are distinct mod~$L$. If all codewords in a CAC $\C$ are equi-difference, then we say that $\C$ is equi-difference, and the set of generators is denoted by $\Gamma(\C)$.

We adopt the terminology in~\cite{Momihara07} and say that a codeword $\set{I}$ of weight $w$ is {\em exceptional} if
\begin{equation}
 |d^*(\set{I})| < 2w-2,  \label{eq:exceptional1}
\end{equation}
or equivalently, if
\begin{equation}
 |d(\set{I})| \leq 2w-2.  \label{eq:exceptional2}
\end{equation}
From the discussion above, we see that if a codeword $\set{I}$ is equi-difference with generator  $g$, then it is exceptional if and only if $\pm g,$ $\pm 2g, \ldots, \pm (w-1)g$ are {\em not} distinct mod $L$.

The CAC in Example~1 is equi-difference, with generators 1, 5, 6 and~11. The codeword generated by 5 is exceptional, because
\[|d^*(\{0,5,10\})| = |\{5,10\}| = 2 < 2\cdot 3-2.
\]

In Example~2, the codewords $\{0,1,2,3\}$, $\{0,4,8,12\}$ and $\{0,5,10,15\}$ are equi-difference, generated by 1, 4, and 5, respectively. The codeword $\{0,6,13,19\}$ is not equi-difference, but it is exceptional, because
\[|d^*(\{0,6,13,19\})| = |\{6,7,13,19,20\}| = 5 < 2\cdot 4-2.
\]
We see that an exceptional codeword is not necessarily equi-difference.

\section{Existing Constructions of CAC in the Literature} \label{sec:known_construction}

The following three constructions of CAC  are due to~\cite{Momihara07}. We state them in this section for the convenience of the readers. The optimality of these constructions is known only for some special cases.
We will show later in Section~\ref{sec:optimal} that they are indeed optimal under more general conditions.

The first and second constructions are based on the multiplicative structure of finite field:
given a prime $p$, the set of non-zero elements in $\mathbb{Z}_p$, denoted by $\mathbb{Z}_p^*$, is a cyclic group with order $p-1$ under multiplication. For a divisor $f$ of $p-1$, we denote the multiplicative subgroup in  $\mathbb{Z}_p^*$ of index $f$ by
\[
\set{H}_0^f(p):=\{x\in \mathbb{Z}_p^* :\, x^{(p-1)/f} \equiv 1 \bmod p \},
\]
and its cosets in the multiplicative subgroup $\mathbb{Z}_p^*$ by $\set{H}_j^f(p)$, for $j=1,\ldots, f-1$.
A set of $f$ elements $\{i_0, i_1,\ldots, i_{f-1}\}$ in $\mathbb{Z}_p^*$ is said to form a {\em system of distinct representatives} of $\{H_j^f(p):\, j=0,1,\ldots, f-1\}$ if each coset $\set{H}_j^f(p)$ contains exactly one element in $\{i_0,\ldots, i_{f-1}\}$.

\begin{construction}[{\cite[Thm 3.1]{Momihara07}} ]
Let $p=2(w-1)m+1$ be a prime number and suppose that $\{1,2,\ldots, w-1\}$ forms a system of distinct representatives of  $\{\set{H}_j^{w-1}(p):\,j=0,\ldots, w-2\}$. Let $\alpha$ be a primitive element in the finite field $\mathbb{Z}_p$ and let $g = \alpha^{w-1}$.
Then the $m$ codewords of weight $w$ generated by $1$, $g$, $g^2, \ldots, g^{m-1}$ form an equi-difference $(2(w-1)m+1,w)$-$\CAC$.
\label{construction_p}
\end{construction}

{\em Example 3:} Let $w = 6$, and $p=421$. 2 is a primitive element in the finite field $\mathbb{Z}_{421}$. We can check that
\begin{eqnarray*}
 1 \equiv &2^{420} &\equiv 2^{84\cdot 5+0} \bmod 421 \\
 2 \equiv &2^{1} &\equiv 2^{0\cdot 5+1}\bmod 421 \\
 3 \equiv &2^{404} & \equiv 2^{80\cdot 5+4} \bmod 421 \\
 4 \equiv &2^{2} &\equiv 2^{0\cdot 5+2}\bmod 421 \\
 5 \equiv &2^{278} &\equiv 2^{55\cdot 5+3}\bmod 421.
\end{eqnarray*}

Hence,
$\{1,2,3,4,5\}$ forms a system of distinct representatives of $\mathcal{H}_0^5(421)$, $\mathcal{H}_1^5(421), \ldots, \mathcal{H}_4^5(421)$. The 42 codewords generated by $2^{5j}$ mod $421$, $j=0,1,\ldots, 41$, form a $(421,6)$-$\CAC$. The generators are:  1, 29, 32, 52, 75, 86, 93, 95, 111, 115,  122, 137, 149, 170, 171, 174, 178, 182, 184,   188,   202,   205,   207,   223,   226,   229,   245,   262,   269,   286,   295,   301,   309,   311,   312,   351, 370,   385,   388,   400,   401, and   415.

\medskip

\begin{construction}[{\cite[Thm 3.7]{Momihara07}} ]
Let $p$ be a prime that can be written as $p=2fm+1$ for some integers $f\geq 1$ and $m\geq 1$. If $s \geq 2$ is an integer such that each of
$\{\pm s, \pm 2s, \ldots, \pm fs\}$ and
\[ \{i+js:\, j=-f, -f+1, \ldots, f-1\}
\]
for $i=1,2,\ldots, s-1$, forms a system of distinct representatives of the cosets of $\{\set{H}_j^{2f}(p):\, j=0,\ldots, 2f-1\}$, then there exists an equi-difference $(s(2fm+1),sf+1)$-$\CAC$ with $m$ codewords. Furthermore, the codewords $\set{I}_1,\ldots, \set{I}_{m}$ satisfies
\[
 \mathbb{Z}_{sp} \setminus \bigcup_{j=1}^m d^*(\set{I}_j) = p \mathbb{Z}_{sp},
\]
where $\alpha \mathbb{Z}_L$ represents the set of integral multiples of $\alpha$ in $\mathbb{Z}_L$.

\label{construction2}
\end{construction}

{\em Example 4:} Consider $w = 7$, $f=2$ and $s=3$. The prime number $p=37$ satisfies the conditions in Construction~\ref{construction2}.
We have
\begin{align*}
\mathcal{H}_0^4(37) &= \{1, 7, 9, 10, 12, 16, 26, 33, 34\},  \\
\mathcal{H}_1^4(37) &= \{2, 14, 15, 18, 20, 24, 29, 31, 32\},  \\
\mathcal{H}_2^4(37) &= \{3, 4, 11, 21, 25, 27, 28, 30, 36\},  \\
\mathcal{H}_3^4(37) &= \{5,6,8,13,17,19,22,23,35 \}.
\end{align*}
We can verify that each of
\begin{align*}
\{\pm 3, \pm 6\} &= \{3,6,31,34\}, \\
\{-5, -2, 1, 4\} &= \{1,4, 32, 35\}, \\
\{-4, -1, 2, 5\} &=\{2,5,33,36\},
\end{align*}
forms a system of distinct representatives of $\mathcal{H}_0^4(37)$,  $\mathcal{H}_1^4(37)$, $\mathcal{H}_2^4(37)$,  $\mathcal{H}_3^4(37)$. By Construction~\ref{construction2}, we have a $(111,7)$-$\CAC$ consisting of $m=9$ codewords. Indeed, the generators of this CAC are 1, 7, 10, 16, 34, 46, 39, 70, 100.

\medskip

The last construction we discuss in this section is a recursive construction.

\begin{construction}[{\cite[Thm 6.1]{Momihara07}} ]
Let $w \geq 3$, and $L_1$, $L_2$ and $s$ be positive integers such that $L_1$ is divisible by $s$ and $\gcd(\ell,L_2) = 1$  for $\ell=2,\ldots, w-1$.
Let $\C_1$ be an equi-difference $(L_1,w)$-$\CAC$ consisting of $m_1$ non-exceptional codewords $\set{I}_1,\ldots, \set{I}_{m_1}$ so that
\[
  \mathbb{Z}_{L_1} \setminus \bigcup_{j=1}^{m_1} d^*(\set{I}_j) \supseteq
  (L_1/s) \mathbb{Z}_{L_1}.
\]
Let $\C_2$ be an equi-difference $(sL_2,w)$-$\CAC$ with $m_2$ codewords. The code $\C$ with length $L_1 L_2$ generated by
$i+jL_1$, for $i \in \Gamma(\C_1), j=0,1,\ldots, L_2-1$, and
$(L_1/s) k$, for $k\in \Gamma(\C_2)$
is an equi-difference $(L_1 L_2, w)$-$\CAC$ with $m_1 L_2 + m_2$ codewords.
 \label{construction_r}
\end{construction}

{\em Example 5:}
The prime numbers $p=37$ and $p=53$ satisfy the conditions in Construction~\ref{construction2} with $w = 7$, $f=2$ and $s=3$. We have a $(3\cdot 37, 7)$-$\CAC$ consisting of $(37-1)/4 =9$ codewords, and a
$(3\cdot 53, 7)$-$\CAC$ consisting of $(53-1)/4 =13$ codewords.
Using Construction~\ref{construction_r} with $L_1 = 3\cdot 53$, $L_2 = 37$, $s=3$ and $w=7$, we obtain a $(3\cdot 37\cdot 53, 7)$-$\CAC$ with $13\cdot 37+9 = 490$ codewords.

\section{Upper Bound on the Size of CAC} \label{sec:Kneser}

In this section we derive an upper bound on the size of CAC.
A tool that we will use is Kneser's theorem~\cite{Kneser53}, which is a result about the sum of subsets in an abelian group~$G$. As we only work with $\mathbb{Z}_L$, we will state  Kneser's theorem for $G = \mathbb{Z}_L$. First we introduce some more notations.

Given two non-empty subsets $\set{A}$ and $\set{B}$ of $\mathbb{Z}_L$, the {\em sum set} and {\em difference set} of $\set{A}$ and $\set{B}$, are defined as
\begin{align*}
 \set{A}+\set{B} &:= \{a+b:\, a\in \set{A}, b \in \set{B}\} \\
 \set{A}-\set{B} &:= \{a-b:\, a\in \set{A}, b \in \set{B}\}
\end{align*}
respectively.
Thus, $\set{I} - \set{I}$ is just another expression for $d(\set{I})$.
We also write $a+\set{B} := \{a\}+ \set{B}$, for non-empty subset $\set{B} \subseteq \mathbb{Z}_L$ and $a\in\mathbb{Z}_L$. The negative of $\set{A}$ is defined as
\[
 -\set{A} := \{-a:\, a \in \set{A} \}.
\]
Given a non-empty subset $\set{S} \subseteq \mathbb{Z}_L$, an element $h \in \mathbb{Z}_L$ is called a {\em period} of $\set{S}$ if $h + \set{S} = \set{S}$. The {\em stabilizer} of $\set{S}$, denoted by $H(\set{S})$, is the set of all periods of~$\set{S}$,
\[
 H(\set{S}) := \{h\in \mathbb{Z}_L:\, h+\set{S}=\set{S}\}.
\]
We note that $0\in H(\set{S})$ for every non-empty subset~$\set{S}$ of $\mathbb{Z}_L$, and $H(\set{S})$ is a subgroup of $\mathbb{Z}_L$. A subset $\set{S}$ is called {\em periodic} if it is non-empty and $H(\set{S}) \neq \{0\}$. If $\set{S}$ is periodic with stabilizer $H$, then  we say that $\set{S}$ is {\em $H$-periodic}.  In other words, a subset of $\mathbb{Z}_L$ is periodic if its stabilizer is a non-trivial subgroup of $\mathbb{Z}_L$.

\begin{lemma}
For any subset $\set{I} \in \mathbb{Z}_L$, we have $d(\set{I}) \supseteq H(d(\set{I}))$.\label{lemma:1}
\end{lemma}

\begin{proof}
Let $h$ be an element in $H(d(\set{I}))$. Because $0 \in d(\set{I})$ and $h+d(\set{I}) \subseteq d(\set{I})$, we have $h = h+0 \in d(\set{I})$. This proves that the stabilizer of $d(\set{I})$ is a subset of $d(\set{I})$.
\end{proof}

Note that an $H$-periodic subset $\mathcal{S}$ of $\mathbb{Z}_L$ can be written as the union of cosets of $H$,
\[ \set{S} = \bigcup_{a\in \mathcal{S}} (H+a).
\]
Conversely, any union of cosets of a non-trivial subgroup~$H$ of $\mathbb{Z}_L$ is $H$-periodic.

We use $\langle \alpha \rangle$ to represent the subgroup of $\mathbb{Z}_L$ generated by~$\alpha$, i.e.,
\[
\langle \alpha \rangle := \{j\alpha \in \mathbb{Z}_L:\, j=0,1,2,\ldots \}.
\]
If $\alpha$ divides $L$, then $\langle \alpha \rangle$ consists of $L/\alpha$ elements.

As an example, consider the subset $\set{S}=\{0,1,3,4\} \subset \mathbb{Z}_6$. The stabilizer of $\set{S}$ is $H = \{0,3\} = \langle 3 \rangle$, and hence  $\set{S}$ is $\langle 3 \rangle$-periodic. We can see that $\set{S}$ is a union of $H$ and the coset $\{1,4\}$.

\begin{theorem}[Kneser]
Let $\set{A}$ and $\set{B}$ be non-empty subsets of~$\mathbb{Z}_L$, and let $H = H(\set{A}+\set{B})$ be the stabilizer of $\set{A}+\set{B}$.
If $|\set{A}+\set{B}| < |\set{A}| + |\set{B}|$, then
\begin{equation}
|\set{A}+\set{B}| = |\set{A}+H| + |\set{B}+H| - |H|. \label{eq:Kneser0}
\end{equation}  \label{thm:Kneser0}
\end{theorem}

The set $\set{A}+H$ can be considered as the ``completion'' of $\set{A}$ with respective to $H$, because $\set{A}+H$ is the smallest $H$-periodic subset in $\mathbb{Z}_L$ which contains~$\set{A}$. Similarly, $\set{B}+H$ can be considered as the completion of $\set{B}$ with respect to~$H$. The conclusion in Kneser's theorem can be rephrased in words as: the cardinality of the sum set of $\mathcal{A}$ and $\mathcal{B}$ is equal to the cardinality of the completion of $\mathcal{A}$ with respective to the stabilizer $H$, plus the cardinality of the completion of $\mathcal{B}$ with respective to the stabilizer $H$,  minus the size of $H$.

Proof of Theorem~\ref{thm:Kneser0} can be found in~\cite{Mann65} or~\cite{Nathanson}. We will apply Kneser's theorem through the following corollary.

\begin{corollary}
Let $\set{I}$ be an exceptional codeword in an $(L,w)$-$\CAC$ and $H$ be the stabilizer of $d(\set{I})$, then $d(\set{I})$ is periodic, and
\begin{equation}
  |d(\set{I})| = 2|\set{I}+H| - |H|. \label{eq:Kneser2}
\end{equation}
\label{cor:Kneser}
\end{corollary}

\begin{proof}
Suppose that $\set{I}$ is an exceptional codeword in an $(L,w)$-$\CAC$ and let $H$ be the stabilizer of $d(\set{I})$.
The condition in Kneser's theorem is satisfied with $\set{A} = \set{I}$ and $\set{B} = -\set{I}$, because
\begin{equation}
|\set{I} + (-\set{I})| = | d(\set{I})| \leq 2w -2 < 2|\set{I}|. \label{eq:G}
\end{equation}
From~\eqref{eq:Kneser0}, we obtain
\begin{align*}
|d(\set{I})| &= |\set{I}+H| + |-\set{I}+H| - |H|   \\
&= |\set{I}+H| + |\set{I}-H| - |H|   \\
 &= 2|\set{I}+H| - |H|.
\end{align*}
In the last equality above, we have used the fact that $H$ is an additive subgroup of $\mathbb{Z}_L$ and hence $-H = H$.
This proves~\eqref{eq:Kneser2}. Since $|\set{I}+H| \geq w$, we obtain
\begin{equation}
|d(\set{I})|  \geq 2w - |H|. \label{eq:H}
\end{equation}
Putting \eqref{eq:G} and \eqref{eq:H} together, we have
\[
2w-|H| \leq  |d(\set{I})| < 2w-1.
\]
We conclude that $|H|>1$ and therefore $d(\set{I})$ is periodic.
\end{proof}

We illustrate Kneser's theorem and Corollary~\ref{cor:Kneser} using Example 1 and~2. In Example~1, consider the exceptional codeword
\[\set{I}_1 = \{0,5,10\} \subset \mathbb{Z}_{15}.
 \]
The stabilizer of $d(\set{I}_1) = \{0,5,10\}$, which is just equal to $\set{I}_1$ itself, is $\langle 5 \rangle$-periodic. We can verify that
\[
|d(\set{I}_1)| = 2|\set{I}_1+\langle 5 \rangle|  - |\langle 5 \rangle| = 2\cdot 3 - 3 = 3.
\]
The codeword $\set{I}_2 =\{0,1,2\}$ in Example 1 is equi-difference and non-exceptional. The condition in Kneser's theorem is satisfied with $\set{A} = \set{I}_2$ and $\set{B} = -\set{I}_2$, since
\[
  |d(\set{I}_2)| = |\set{I}_2 - \set{I}_2| =|\{0,\pm1, \pm2\}| = 5 < 2|\set{I}_2|.
\]
We have $H(d(\set{I}_2)) = \{0\}$, and
\[
 |d(\set{I}_2)| = |\set{I}_2+\{0\}| + |\set{I}_2-\{0\}| - |\{0\}| = 5.
\]

In Example 2, consider the exceptional codeword
\[\set{I} = \{0,6,13,19\}\subset\mathbb{Z}_{26}.
\]
The corresponding set of differences
\[
 d(\{0,6,13,19\}) = \{0,6,7, 13, 19, 20\}
\]
is $\langle 13 \rangle$-periodic. We can check that
\[
|d(\set{I})| = 2|\set{I}+\langle 13 \rangle|  - |\langle 13 \rangle| = 2\cdot 4 - 2 = 6.
\]

The next theorem provides a recipe for upper bounding the size of a CAC.

\begin{theorem}
Let $\C$ be an $(L,w)$-$\CAC$ in which $E$ codewords are exceptional.  For $j=1,2,\ldots, E$, denote the $j$-th exceptional codeword by $\set{I}_j$, and let the stabilizer of $d(\set{I}_j)$ be~$H_j$.
Define
\begin{equation}
\Delta_j := |\set{I}_j + H_{j}| - w.
\label{eq:def_Delta}
\end{equation}
Then
\begin{equation}
|\C| \leq  \frac{L-1+ \sum_{j=1}^E (|H_j|-1 - 2\Delta_j ) }{2w-2} . \label{eq:lemma}
\end{equation}
\label{lemma}
\end{theorem}

\begin{proof}
By definition, $d^*(\set{I})$ and $d^*(\set{J})$ are disjoint for any pair of  distinct codewords $\set{I}$ and $\set{J}$ in~$\C$. We have the following basic inequality,
\begin{equation}
 L-1  \geq \sum_{\set{I}\in\C} |d^*(\set{I})|. \label{eq:basic0}
\end{equation}
Let the number of non-exceptional codewords be~$N$. Since $d^*(\set{I}) \geq 2w-2$ for each non-exceptional codeword $\set{I}$, the inequality in \eqref{eq:basic0} becomes
\[
 L-1 \geq N(2w-2) + \sum_{j=1}^E |d^*(\set{I}_j)|.
\]
From Corollary~\ref{cor:Kneser} we get
\begin{align*}
 \sum_{j=1}^E |d^*(\set{I}_j)| &= \sum_{j=1}^E \Big(|d(\set{I}_j)| -1 \Big) \\
 & =  \sum_{j=1}^E \Big( 2|\set{I}_j+H_j| - |H_j| -1\Big).
\end{align*}
Therefore,
\begin{align*}
 L-1 & \geq N(2w-2) + \sum_{j=1}^E \Big(2|\set{I}_j + H_j| -|H_j| -1 \Big) \\
 & = (N+E)(2w-2) +  \sum_{j=1}^E \Big(2\Delta_j -|H_j| +1 \Big).
\end{align*}
After some rearrangement of terms, we get
\[
 |\C| = N+E \leq  \frac{L-1 +  \sum_{j=1}^E (|H_j|-1-2\Delta_j)}{2w-2} .
\]
This finishes the proof of the theorem.
\end{proof}

We note that the value of $\Delta_j$ defined in~\eqref{eq:def_Delta} is non-negative for all $j$, because
$\Delta_j = |\set{I}_j + H_j| - |\set{I}_j|$, and $\set{I}_j$ is a subset of
$\set{I}_j + H_j$. We have the following corollary.

\begin{corollary}
Let $\C$ be an $(L,w)$-$\CAC$. If there are $E$ exceptional codewords $\set{I}_1$, $\set{I}_2,\ldots, \set{I}_E$, in $\C$, then
\begin{equation}
|\C| \leq  \frac{L-1+ \sum_{j=1}^E (|H(d(\set{I}_j))|-1) }{2w-2}.
\end{equation}
\label{cor:weak_upper_bound}
\end{corollary}

We make a few more definitions. The motivation of these definition will be clear after Theorem~\ref{thm:main1}. Let
\begin{align}
S(L,w) := \Big\{x &\in \{2,3,\ldots, 2w-2\}:\,
x \text{ divides } L, \label{eq:defS1.5} \\
& \qquad  \text{ and } 2x \lceil w/x \rceil - x \leq 2w-2 \Big\}. \label{eq:def_S2}
\end{align}
$S(L,w)$ may be empty, for example when $L$ is prime.
Let $\mathscr{S}(L,w)$ be the collection of subsets of $S(L,w)$, such that each pair of distinct elements in $\set{S} \in \mathscr{S}(L,w)$ are relatively prime, i.e.,
\[
 \mathscr{S}(L,w) := \{\set{S} \subseteq S(L,w) :\, \gcd(i,j)=1, \forall i,j,\in \set{S}, i\neq j\}.
\]

Given an integer $L \geq w \geq 2$, if $\mathscr{S}(L,w)$ is non-empty, define
\begin{equation}
 F(L,w) := \max_{\mathcal{S} \in \mathscr{S}(L,w)}  \sum_{x \in \mathcal{S}} \Big(x - 1 - 2x \lceil w/x \rceil  + 2w \Big)  \label{eq:def_F}
\end{equation}
with the maximum taken over all subsets $\mathcal{S}$ in $\mathscr{S}(L,w)$.  If $\mathscr{S}(L,w)$ is empty, we define $F(L,w)$ as zero.  We note that the summand in~\eqref{eq:def_F} is positive by the condition in~\eqref{eq:def_S2}. Hence,  $F(L,w)$ is non-negative.

\begin{theorem}
For $L \geq w \geq 2$,
\begin{equation}
 M(L,w) \leq \left\lfloor \frac{L-1+F(L,w)}{2w-2} \right\rfloor.
 \label{eq:main1}
\end{equation}
\label{thm:main1}
\end{theorem}

\begin{proof}
Let $\C$ be an $(L,w)$-$\CAC$. If there is no exceptional codeword in $\C$, then $|\C| \leq \lfloor (L-1)/(2w-2) \rfloor$ by Theorem~\ref{lemma}. Since $F(L,w)$ is non-negative, the size of $\C$ is less than or equal to the right hand side of~\eqref{eq:main1}.

Suppose that there are $E$ exceptional codewords in an $(L,w)$-$\CAC$, denoted by $\set{I}_1$, $\set{I}_2, \ldots, \set{I}_E$.  For $j=1,2,\ldots, E$, let $H_j$ be the stabilizer of $d(\set{I}_j)$.
Let $i\neq j$ and consider two distinct exceptional codewords $\set{I}_j$ and $\set{I}_j$ in~$\C$. Both $|H_i|$ and $|H_j|$ are strictly larger than one by Corollary~\ref{cor:Kneser}. We claim that $|H_i|$ and $|H_j|$ are relatively prime. As subgroups of~$\mathbb{Z}_L$, $H_i$ and $H_j$ can be written as $\langle \alpha_i \rangle$ and $\langle \alpha_j \rangle$ respectively, for some proper divisors $\alpha_i$ and $\alpha_j$ of~$L$, so that $|H_i|=L/\alpha_i$ and $|H_j| = L/\alpha_j$. If $|H_i|$ and $|H_j|$ are not relatively prime, say, if $b > 1$ is a common divisor of $|H_i|$ and $|H_j|$, then
\[
 b x_i = \frac{L}{\alpha_i}, \  b x_j = \frac{L}{\alpha_j},
\]
for some integers $x_i$ and $x_j$, and we get
\[
 b \alpha_i x_i = L = b \alpha_j x_j.
\]
After dividing the above equation by $b$, we see that $L/b$ is an integral multiple of both $\alpha_i$ and  $\alpha_j$, and hence is a common element in $H_i$ and $H_j$. Moreover,  $L/b$ is non-zero mod $L$, because $b > 1$. The two stabilizers $H_1$ and $H_2$ thus contain a common non-zero element. By Lemma~\ref{lemma:1}, we have $d(\set{I}_i) \supseteq H_i$ and $d(\set{I}_j) \supseteq H_j$, and so $L/b$ is also a common non-zero element of $d(\set{I}_i)$ and $d(\set{I}_j)$.
This contradicts the defining property that $d(\set{I}_i)\cap d(\set{I}_j)=\{0\}$. This completes the proof of the claim.

For each $j$, $|\set{I}_j+H_j|$ is an integral multiple of $|H_j|$ because
$\set{I}_j+H_j$ is a union of $H_j$ and its cosets. Furthermore, as we have already noted in the proof of Corollary~\ref{cor:Kneser}, $|\set{I}_j+H_j|$ is larger than or equal to $w$ because $\set{I}_j+H_j$ contains $\set{I}_j$. We thus have the following inequality,
\[
 |\set{I}_j+H_j| \geq |H_j| \left\lceil \frac{w}{|H_j|} \right\rceil .
\]
The right hand side in the above inequality is the smallest integral multiples of $|H_j|$ which is larger than or equal to $w$.

We next show that $|H_j| \in S(L,w)$, for $j=1,2,\ldots, E$. For each $j$,
the subgroup $H_j$ cannot have size strictly larger than $2w-2$, otherwise by Corollary~\ref{cor:Kneser}, we have
\begin{align*}
|d(\set{I}_j)| & = 2|\set{I}_j+H_j|  - |H_j| \notag \\
& \geq 2|H_j|  - |H_j| \\
& = |H_j| > 2w-2,
\end{align*}
which is a contradiction to the definition of exceptional codeword in~\eqref{eq:exceptional2}. In addition, we must have $|H_j| \geq 2$ because $\set{I}_j$ is  periodic by assumption.
This shows that $2\leq |H_j| \leq 2w-2$.

As a subgroup of $\mathbb{Z}_L$, we see that $|H_j|$ is a divisor of~$L$.
Moreover, for $j=1,2,\ldots, E$, $|H_j|$ satisfies
\begin{align*}
2w-2 \geq d(\set{I}_j) &= 2  |\set{I}_j+H_j| - |H_j| \notag \\
& \geq 2 |H_j|  \left\lceil \frac{w}{|H_j|} \right\rceil - |H_j|.
\end{align*}
Consequently, $|H_j|$ satisfies the conditions in \eqref{eq:defS1.5} and~\eqref{eq:def_S2}, and hence belong to the set $S(L,w)$. We have already shown that $|H_i|$ and $|H_j|$ are relatively prime for $i\neq j$. Therefore
\[
 \{|H_1|, |H_2|, \ldots, |H_E|\} \in \mathscr{S}(L,w).
\]

For $j=1,2,\ldots, E$, let $\Delta_j$ be defined as in Theorem~\ref{lemma}. We can upper bound  $|H_j|-1-2 \Delta_j$, which appears in the summation in~\eqref{eq:lemma}, by
\[
 |H_j| - 1 - 2 \Delta_j
 \leq |H_j| - 1 - 2|H_j| \left\lceil \frac{w}{|H_j|} \right\rceil  + 2w,
\]
which equals the summand in~\eqref{eq:def_F} with $x$ substituted by $|H_j|$.
By exhausting all possible choices of $\set{S}$ in $\mathscr{S}(L,w)$, we have the following upper bound
\[
 \sum_{j=1}^E (|H_j|-1-2 \Delta_j) \leq F(L,w).
\]
Substituting it back to~\eqref{eq:lemma}, we have
\[
| \C|
\leq  \left\lfloor \frac{L-1+F(L,w)}{2w-2} \right\rfloor
\]
This completes the proof of Theorem~\ref{thm:main1}.
\end{proof}

For CAC with weight $w=3$ and odd length $L$, we can check that $S(L,3)$ is either empty or $\{3\}$. So in the computation of $F(L,3)$ in~\eqref{eq:def_F}, the maximum is taken over only one number, namely $x=3$, and we get
$$F(L,3) = 3 - 1 - 2\cdot 3 \lceil 3/3 \rceil + 2\cdot 3 = 2.$$
Hence from Theorem~\ref{thm:main1}, we obtain
$$ M(L,3) \leq \left\lfloor \frac{L+1}{4} \right\rfloor.$$
It can be shown that the above bound holds for even length $L$ as well.
This yields the upper bound on the size of CAC for three active users in~\cite{LT05}. When $w=4$ and $w=5$, the upper bounds obtained from Theorem~\ref{thm:main1} coincides with the known results in~\cite[Lemma 2.1, 2.3]{Momihara07}.  We illustrate Theorem~\ref{thm:main1} with $w=6$.

\begin{corollary}
Let $L$ be an integer factorized as $2^p 3^q 7^r \ell$, where $\ell$ is not divisible by 2, 3 or 7. Then we have
\[
 M(L,6) \leq \begin{cases}
 \lfloor (L-1)/10 \rfloor & \text{ if } p=q=r=0, \\
 \lfloor L/10 \rfloor & \text{ if } 1\leq p\leq 2, q=r=0, \\
 \lfloor (L+1)/10 \rfloor & \text{ if } q>1, p=r=0, \\
 \lfloor (L+2)/10 \rfloor & \text{ if } p\geq 3, q=r=0, \\
 \lfloor (L+3)/10 \rfloor & \text{ if } p=q=0,r\geq 1, \\
 \lfloor (L+4)/10 \rfloor & \text{ if } p\geq 1, q\geq 1,r=0, or\\
 & \text{\ \ } 1\leq p \leq 2, q=0, r\geq 1, \\
 \lfloor (L+5)/10 \rfloor & \text{ if } p=0, q\geq 1,r\geq 1, \\
 \lfloor (L+6)/10 \rfloor & \text{ if } p\geq 3,q=0, r\geq 1, \\
 \lfloor (L+8)/10 \rfloor & \text{ if } p\geq 1,q\geq 1, r \geq 1. \\
 \end{cases}
\]
\label{cor:w6}
\end{corollary}

\begin{proof}
The value of $x-1 -2x\lceil w/x \rceil + 2w$ for $x\in\{2,3,\ldots,10\} \setminus \{4,5\}$ is shown in the following table:
\[
\begin{array}{|c|ccccccc|} \hline
x                               &2&3&6&7&8&9&10 \\ \hline
x-1 -2x\lceil w/x \rceil + 2w &1&2&5&4&3&2&1 \\ \hline
\end{array}
\]
We note that 4 and 5 are not shown in the above table, because they do not satisfy the condition in~\eqref{eq:def_S2}.

Since the value of $x-1 -2x\lceil w/x \rceil + 2w$ for $x=2$ and $x=10$ are the same, we can disregard the case $x=10$ in the computation of $F(L,w)$ without affecting the result. We tabulate $S(L,6)$ and $F(L,6)$ in Table~\ref{table:w6}. By Theorem~\ref{thm:main1}, we get
\[M(L,6) \leq \left\lfloor  \frac{L-1+F(L,6)}{10} \right\rfloor.
\]
The upper bound in Corollary~\ref{cor:w6} is obtained after tidying up the data in Table~\ref{table:w6}.
\end{proof}

\begin{table}
\[
\begin{array}{|c|c|c||c|c|} \hline
 p & q & r  & S(2^p 3^q 7^r\ell,6) & F(2^p 3^q 7^r\ell,6)\\ \hline \hline
0 & 0  & 0         & \emptyset     &  0 \\
1,2  &    0 & 0    & \{2\}         & 1 \\
\geq3&0     & 0    & \{2,8\}       & 3\\
0    & 1    & 0    & \{3\}         & 2 \\
1,2  & 1    & 0    & \{2,3,6\}     & 5 \\
\geq3&1     & 0    & \{2,8,3,6\}   & 5 \\
0    &\geq2 & 0    & \{3,9\}       & 2 \\
1,2  & \geq2& 0    & \{2,3,6,9\}       & 5 \\
\geq3& \geq2& 0    & \{2,3,6,8,9\} & 5 \\
0 & 0  & \geq1      & \{7\}     &  4 \\
1,2  &    0 &  \geq1   & \{2,7\}         & 5 \\
\geq3&0     &  \geq1   & \{2,7,8\}       & 7\\
0    & 1    &  \geq1   & \{3,7\}         & 6 \\
1,2  & 1    &  \geq1   & \{2,3,6,7\}     & 9 \\
\geq3&1     &  \geq1   & \{2,3,6,7,8\}   & 9 \\
0    &\geq2 &  \geq1   & \{3,7,9\}       & 6 \\
1,2  & \geq2&  \geq1   & \{2,3,6,7,9\}   & 9 \\
\geq3& \geq2&  \geq1   & \{2,3,6,7,8,9\} & 9 \\ \hline
\end{array}
\]
\caption{Values of $S(L,6)$ and $F(L,6)$}
\label{table:w6}
\end{table}

{\bf Remark:} The value of $F(L,w)$ in Theorem~\ref{thm:main1} can be computed by linear programming as follows. For each element~$i$ in $S(L,w)$, define a variable $z_i$. Let the objective function be $\sum_{i\in S(L,w)} c_i z_i$, with $c_i$ defined by
\[
 c_i :=  i - 1 -2i \lceil w/i \rceil + 2w. \label{eq:objective}
\]
For each prime number $p$ between 2 and $2w-2$, impose a constraint
\begin{equation}
 \sum_{p|i} z_i \leq 1, \label{eq:constraint}
\end{equation}
where the summation is taken over all $i$ that is divisible by~$p$. Then $F(L,w)$ is the optimal solution if we maximize $\sum_{i\in S(L,w)} c_i z_i$ subjective to the constraint in~\eqref{eq:constraint} for $p$ ranging over all prime numbers between 2 and $2w-2$, and $0 \leq z_i \leq 1$ for all $i\in S(L,w)$.

\medskip

\begin{figure}
\begin{center}
  \includegraphics[width=3.7in]{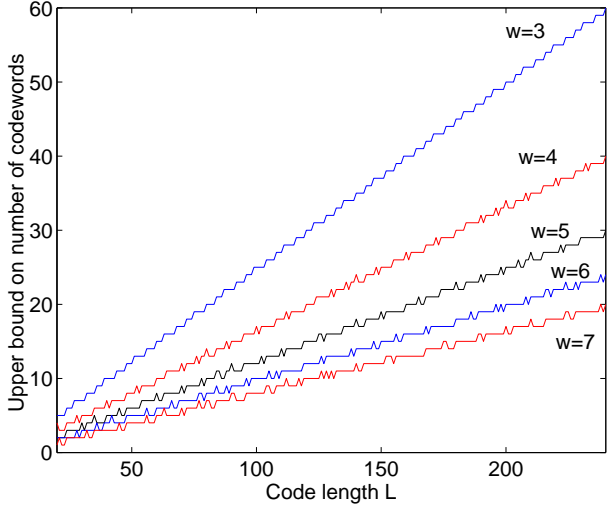}
\end{center}
\caption{Upper bounds on size of CAC for weight 3 to 7.}
\label{fig:graph}
\end{figure}

Using the linear programming mentioned in the above remark, the upper bounds given by Theorem~\ref{thm:main1} for weight 3 to 7 and length between 20 and 240 are plotted in Fig.~\ref{fig:graph}. The lines corresponding to $w=4$ and $w=5$ are the same as the upper bounds on the size of CAC in~\cite{Momihara07}. For each $w$, the growth is roughly linear in $L$, with slope  $(2w-2)^{-1}$. We note that for $w > 3$, the upper bounds are {\em not} monotonically increasing with~$L$.

The computation of $F(L,w)$ amounts to solving a linear programming, and it is not obvious from~\eqref{eq:def_F} how to get an estimate on the value of $F(L,w)$. The next theorem gives an upper bound on $F(L,w)$ in closed-form expression, from which we can analyze the asymptotic growth rate of $M(L,w)$.

Given a positive integer $x \geq 2$, let $\pi(x)$ denote the number of distinct prime numbers between 2 and $x$,
\[
 \pi(x) := |\{i:\, 2\leq i \leq x, i \text{ is prime} \}|.
\]
Note that $\pi(x)$ also counts the maximum number of relatively prime integers between 2 and~$x$.

\begin{theorem}
For $L \geq w \geq 2$,
\begin{equation}
M(L,w) \leq \left\lfloor \frac{L-1}{2w-2} + \frac{\pi(2w-2)}{2} \right\rfloor.
\label{eq:general_bound}
\end{equation}
\label{thm:general_bound}
\end{theorem}

\begin{proof}
Recall that  $F(L,w)$ is the maximum of
\begin{equation}
\sum_{x \in \mathcal{S}} (x - 1 - 2x \lceil w/x \rceil  + 2w ),
\label{eq:sum}
\end{equation}
taken over all subsets $\mathcal{S}$ in $\mathscr{S}(L,w)$.
For $w\leq x$, we observe that
\begin{align*}
x - 1 -2x \lceil w/x \rceil + 2w &= x - 1 -2x+  2w \\
&= 2w-x-1 \\
&\leq w-1,
\end{align*}
and for $w > x$, we have
\begin{align*}
x - 1 -2x \lceil w/x \rceil + 2w
&\leq  x - 1 -2x(w/x)+  2w \\
& = x-1 < w-1.
\end{align*}
In summary,  we obtain
$$x - 1 -2x \lceil w/x \rceil + 2w \leq w-1$$
for all $x\in S(L,w)$.

The number of summands in~\eqref{eq:sum} is less than or equal to the maximum number of relatively prime integers in $S(L,w)$. Since $S(L,w) \subseteq \{2,3,\ldots, 2w-2\}$, the number of summands in~\eqref{eq:sum} is less than or equal to the maximal number of relatively prime integers between 2 and $2w-2$, namely $\pi(2w-2)$. The summation in~\eqref{eq:sum} is thus less than or equal to $(w-1) \pi(2w-2)$, and hence
\[
  F(L,w) \leq (w-1) \pi(2w-2).
\]
Theorem~\ref{thm:general_bound} follows by replacing $F(L,w)$ by $(w-1) \pi(2w-2) $ in Theorem~\ref{thm:main1}.
\end{proof}

{\bf Remark:} The celebrated prime number theorem says that $\pi(x) \log(x) /x$ approaches 1 when $x$ approaches infinity. A weaker form of the prime number theorem proved by Chebyshev~\cite{HardyWright} states that for some constants $B_1 < 1$ and $B_2 > 1$, we can bound $\pi(x)$ by
\[
B_1 \frac{x}{\log(x)} < \pi(x) < B_2 \frac{x}{\log(x)},
\]
for all $x$. Furthermore, $\pi(x)$ can be upper bounded by
\[
 \pi(x) < \frac{x}{\log x - 1.5}
\]
for $x \geq 5$~\cite{Rosser}. Hence, for $w \geq 4$ we have
\[
 M(L,w) \leq \left\lfloor \frac{L-1}{2w-2} + \frac{2w-2}{2 \log(2w-2)-3}\right \rfloor.
\]

\section{Optimality of Existing Constructions of CAC} \label{sec:optimal}
For Hamming weight $w=4$ and $w=5$,  Constructions 1 and 2 are shown to be optimal in~\cite{Momihara07}. In this section,
we use the upper bounds on size of CAC given in Section~\ref{sec:Kneser} to show the optimality of some CACs  by Constructions 1, 2, and 3 with general weight.

\begin{theorem}
All CACs produced by Constuction~\ref{construction_p} are optimal. If $p$ and $w$ satisfy the conditions in Construction~\ref{construction_p}, then we have $M(p,w) = (p-1)/(2w-2)$.
\end{theorem}

\begin{proof}
Since $p$ is prime, there is no nontrivial additive subgroup in~$\mathbb{Z}_p$, and hence there is no exceptional codeword. The upper bound in Theorem~\ref{lemma} reduces to $\lfloor (L-1)/(2w-2)\rfloor$. By definition, $(L-1)/(2w-2) = (p-1)/(2w-2) =m$, which equals the number of codewords in Construction~\ref{construction_p}. The number of codewords meets the upper bound and the constructed CAC is therefore optimal.
\end{proof}

\begin{theorem}
If $m > s$, then any CAC with parameters stated in Construction~\ref{construction2} is optimal. In other words, if $p$, $m$, $s$ and $w$ satisfy the conditions in Construction~\ref{construction2} and $m>s$, then $M(sp,w) = s(p-1)/(2w-2)$. \label{thm:direct_construction}
\end{theorem}

\begin{proof}
Since $m>s$ and $p=2fm+1$, we have $s < p$ and hence $s$ and $p$ are relatively prime.

Let $\C$ be a CAC with length $L=sp$ and weight $w=sf+1$. Suppose that there are $E$ exceptional codewords in $\C$, say $\set{I}_1$, $\set{I}_2,\ldots, \set{I}_E$. Let $H_j = \langle \alpha_j \rangle$, for $j=1,2,\ldots, E$, respectively, be the stabilizer of $d(\set{I}_j)$, where $\alpha_j$'s are divisors of $L=sp$.

Suppose that $\gcd(\alpha_j,p)=1$ for some~$j$. Because $p$ and $s$ are relatively prime, $\alpha_j$ must divide $s$, and hence we get $\langle s \rangle \subseteq \langle \alpha_j \rangle$ and
\[
|H_j| = |\langle \alpha_j \rangle| \geq |\langle s \rangle|.
\]
However,
\begin{equation}
 |\langle s \rangle| =p = 2fm+1 > 2fs+1 =  2w-1, \label{eq:optimal2}
\end{equation}
which implies that $|H_j| > 2w-1$,
contradicting the assumption that $\set{I}_j$ is exceptional. (We have used the assumption that $m>s$ in~\eqref{eq:optimal2}.)
Consequently, $\alpha_j$ is divisible by $p$ for all $j = 1,2,\ldots, E$.

We obtain
\[
H_j =\langle \alpha_j \rangle \subseteq \langle p \rangle
\]
for all $j$, and $\cup_{j=1}^E H_j \subseteq \langle p \rangle$. Since $H_i \cap H_j  = \{0\}$ for $i\neq j$, we obtain
\[
\sum_{j=1}^E (|H_j| -1) \leq |\langle p \rangle| = s.
\]

From Corollary~\ref{cor:weak_upper_bound}, we obtain
\begin{align*}
|\C|
&\leq \left\lfloor \frac{L-1+s-1}{2w-2} \right\rfloor \\
&= \left\lfloor\frac{s(2fm+1) - 1  + s -1}{2fs} \right\rfloor\\
&= \left\lfloor m + \frac{s-1}{fs} \right\rfloor
\end{align*}
From $s-1 < fs$, we conclude that $|\C| \leq m$.
\end{proof}

From Theorem~\ref{thm:direct_construction}, we can construct infinitely many optimal CACs for each  $w\geq 3$. The following is an illustration for $w=7$.

\begin{corollary}
Let $p$ be a prime number congruent to 31 or 39  mod~40.
Then $M(6p,7) = (p-1)/2$.\label{cor:7}
\end{corollary}

\begin{proof}
Apply Construction~\ref{construction2} with $f=1$ and $s=6$. We want to find integer $m$ such that $p=2m+1$ is prime, and each of the following
\[
 \{-5, 1\},  \{-4, 2\},  \{-3, 3\},  \{-2, 4\},  \{-1, 5\},  \{-6, 6\},
\]
forms a system of distinct representatives of $\set{H}_0^2(p)$ and $\set{H}_1^2(p)$. Expressed in terms of the  Legendre symbol $\Legendre{\cdot}{p}$, it is equivalent to
\begin{equation}
\Legendre{-1}{p} = -1, \label{eq:QR1}
\end{equation}
and
\begin{equation}
\Legendre{2}{p} = \Legendre{5}{p} = 1. \label{eq:QR2}
\end{equation}
By the law of quadratic reciprocity~\cite{IrelandRosen}, \eqref{eq:QR1} and \eqref{eq:QR2} are equivalent to the following conditions
\[
\begin{cases}
p \equiv 3 \bmod 4 \\
p \equiv \pm 1 \bmod 8 \\
p \equiv 1,4 \bmod 5
\end{cases}
\]
which can be further simplified to $p \equiv 31 \text{ or } 39 \bmod 40$.
Hence, for each prime $p \equiv 31 \text{ or } 39 \bmod 40$, we have a $(6p,7)$-$\CAC$ consisting of $(p-1)/2$ codewords, which is optimal by Theorem~\ref{thm:direct_construction}.
This proves that $M(6p,7) = (p-1)/2$.
\end{proof}

By Dirichlet's theorem on primes in arithmetic progression~\cite{HardyWright}, there are infinitely many prime $p$ that satisfies $p \equiv 31 \text{ or } 39 \bmod 40$. We thereby have infinitely many optimal CACs with weight $w=7$. The argument for $w=7$ can be adopted to all weight~$w$ to construct infinitely many optimal CACs for each~$w$.

Applying the recursive construction in Construction~\ref{construction_r} to
$(6p_i,7)$-$\CAC$, for $i=1,2,\ldots, n$, where $p_i$ is prime and congruent to  31  or  39 mod 40, we obtain $(6p_1p_2\cdots p_n,7)$-$\CAC$.  By similar argument as in the proof of Theorem~\ref{thm:direct_construction}, we can show that the resulting $(6p_1p_2\cdots p_n,7)$-$\CAC$ is optimal. This proves the following corollary.

\begin{corollary}
Let $p_i$ be prime number that satisfies $p_{i} \equiv 31 \text{ or } 39 \bmod 40$, for $i=1,2,\ldots, n$. Then
\[
 M(6p_1p_2\cdots p_n,7) = (p_1p_2\cdots p_n - 1)/2.
\]
\end{corollary}

\medskip

In the remaining of this section, we apply the upper bound in this paper to show that Construction~\ref{construction_r} produces optimal CAC for some special choices of input parameters.

\begin{theorem}
Suppose $p$ is a prime number such that $p-1$ is divisible by $2w-2$ and $p>2w-1$. If there is an equi-difference $(p,w)$-$\CAC$ with $m=(p-1)/(2w-2)$ codewords, then
\[M(p(2w-1),w) = p+m.\]
 \label{thm:extend}
\end{theorem}

\begin{proof}
We apply the recursive Construction~\ref{construction_r} with $s=1$, $L_1=2w-1$, $L_2 = p$, and
take $\C_1$ to be a trivial $(2w-1,w)$-$\CAC$ consisting of $m_1 = 1$ codeword generated by~1, and $\C_2$ to be the given equi-difference $(p,w)$-$\CAC$ with $m_2= m$ codewords. It is implied by the assumption $p > 2w-1$ that $\gcd(p,2w-1)=1$. So the condition $$\gcd(\ell, L_2) = \gcd(\ell,p) = 1$$ is satisfied for $\ell = 2,3,\ldots, w-1$.
Construction~\ref{construction_r} yields a $(p(2w-1),w)$-$\CAC$ with $m_1p+m_2 = p+m$ codewords.

It remains  to show that any $(p(2w-1),w)$-$\CAC$ contains at most $p+m$ codewords.
Let $\C$ be a $(p(2w-1),w)$-$\CAC$. Suppose that $\set{I}_j$, for $j=1,2,\ldots, E$, are the exceptional codewords in~$\C$, and $H_j$ is the stabilizer of~$d(\set{I}_j)$.
For each~$j$, $H_j$ contains strictly less than $2w-1$ elements because $|H_j| \leq |d(\set{I}_j)| \leq 2w-2$.

We claim that any additive subgroup $G$ in $\mathbb{Z}_{p(2w-1)}$ of size strictly less than $2w-1$ is included in
$$\langle p \rangle = \{0, p, 2p, \ldots, (2w-2)p\}.$$
Suppose on the contrary that we can find an $a \in G$ not divisible by $p$. Then $\gcd(a, p(2w-1))$ is a divisor of $2w-1$. The order of $a$ in $G$, which equals
\[
 \frac{p(2w-1)}{\gcd(a,p(2w-1))},
\]
is thus larger than or equal to $p$. Since $p > 2w-1$ by hypothesis, the integral multiples of $a$ in $G$ already generate more than $2w-1$ distinct elements, contradicting the assumption that $|G| < 2w-1$. Therefore, any integer $a$ which is not an integral multiple of $p$ does not belong to $G$.

By the above claim, we have $H_j \subseteq \langle p \rangle$ for each $j$.
We can write $H_j = \langle \beta_j p \rangle$, for some $\beta_j$ between 1 and $2w-2$. However, $\beta_j$ cannot be relatively prime with $2w-1$, otherwise, the integral multiples of $\beta_jp$ would generate $\langle p \rangle$ and we would have  $|H_j| = |\langle p \rangle| = 2w-1$, contradicting $|H_j| < 2w-1$. In particular, we obtain
\[
 H_j \subseteq \{ zp:\, z=0,1,\ldots, 2w-2,\ \gcd(z,2w-1) > 1\}.
\]

Since $H_i \cap H_j = \{0\}$ for $i\neq j$, we get
\begin{align*}
 \sum_{j=1}^E (|H_j| - 1)
& \leq |\langle p \rangle|  -1 - \varphi(2w-1) \\
& \leq (2w-2)-\varphi(2w-1),
\end{align*}
where $\varphi(x)$ denotes the number of integers in $\{1,2,\ldots, x-1\}$ which are relatively prime with~$x$. By Corollary~\ref{cor:weak_upper_bound}, we obtain
\[
 |\C| \leq \left\lfloor \frac{L-1}{2w-2} + \frac{2w-2-\varphi(2w-1)}{2w-2} \right\rfloor.
\]
Since
\[
\frac{L - 1}{2w-2} = \frac{p(2w-1) -1}{2w-2} = p+m,
\]
and
\[
\frac{2w-2-\varphi(2w-1)}{2w-2} < 1
\]
we conclude that $M(p(2w-1),w) = p+m$.
\end{proof}

{\em Example 6:} In~\cite{Momihara07}, CAC of length $p$ and weight 4, containing $(p-1)/6$ codewords, is reported for infinitely many prime number~$p$ using Construction~\ref{construction_p}. We can extend each of them to an optimal $(7p,4)$-$\CAC$ with $p+(p-1)/6$ codewords. The example with smallest $p$ is a $(37,4)$-$\CAC$ with 6 codewords generated by 1, 8, 27, 31, 26, and~23.
It can be lengthened and enlarged to an optimal  $(259,4)$-$\CAC$ with 43 codewords.

\begin{theorem} Let $w \geq 3$, and  $p$ be a prime number satisfying
\begin{equation}
w\leq p \leq w+ \varphi(2w-1)/2,\label{eq:CRT_condition}
\end{equation}
where $\varphi(2w-1)$ denotes the number of integers in $\{1,2,\ldots, 2w-2\}$ which are relatively prime with $2w-1$. We have
\[M(p(2w-1),w) = p+1.\]  \label{thm:CRT}
\end{theorem}

\begin{proof} We apply Construction~\ref{construction_r} with $s=1$, $L_1=2w-1$ and $L_2 = p$.
Let $\C_1$ be a $(2w-1,w)$-$\CAC$ containing only one equi-difference codeword generated by~1. Let $\C_2$ be a $(p,w)$-$\CAC$ containing only one equi-difference codeword generated by~1. Since $w\leq p$ and $p$ is prime, the condition $\gcd(\ell,p) = 1 $ for $\ell=2,3,\ldots, w-1$ is satisfied.
By Construction~\ref{construction_r}, we have a $(p(2w-1), w)$-$\CAC$ containing $p+1$ codewords.

We now show that this is an optimal CAC with length $L=p(2w-1)$ and weight~$w$.
Let $\C$ be any $(p(2w-1), w)$-$\CAC$ with $p$ prime and satisfying the condition in~\eqref{eq:CRT_condition}. We show that $\C$ contains at most $p+1$ codewords by considering the following two cases.

{\it Case 1:}  there is no exceptional codeword in $\C$. By Theorem~\ref{lemma} we obtain
\begin{align}
|\C| & \leq \left\lfloor \frac{L-1}{2w-2} \right\rfloor = \left\lfloor \frac{p(2w-1)-1}{2w-2} \right\rfloor \notag\\
  &= p+ \left\lfloor \frac{p-1}{2w-2} \right\rfloor . \label{eq:CRT_inequality}
\end{align}
However, by the second inequality in~\eqref{eq:CRT_condition},
\[
 p-1 \leq w-1 + \frac{\varphi(2w-1)}{2}.
\]
Since $\varphi(2w-1) \leq 2w-2$, we get $p-1 \leq 2w-2$, and hence $|\C| \leq p+1$ by~\eqref{eq:CRT_inequality}.

{\it Case 2:} there is at least one exceptional codeword in~$\C$.
Let $\set{I}$ be an exceptional codeword in~$\C$.
We first prove the following claim: the stabilizer $H$ of $d(\set{I})$ is either a subset of $\langle p \rangle$ or equal to $\langle 2w-1 \rangle$. Let $H = \langle \alpha \rangle$, where $\alpha$ is a proper divisor of $L=p(2w-1)$. If $\alpha$ is divisible by $p$, then $\langle \alpha \rangle \subseteq \langle p \rangle$, and we have $H \subseteq \langle p \rangle$. Otherwise, if $\alpha$ is not divisible by $p$, then $\alpha$ divides $2w-1$. Suppose that $2w-1$ is factorized as $\alpha \beta$. We have
\[
 |\langle \alpha \rangle| = \frac{p(2w-1)}{\alpha} = p\beta.
\]
If $\alpha$ is strictly less than $2w-1$, then $\beta \geq 2$, and thus
$|\langle \alpha \rangle| \geq 2p$. As $p\geq w$ by assumption, we obtain
\[
 |d(\set{I})| \geq |\langle \alpha \rangle| \geq 2p \geq 2w,
\]
which is a contradiction to the hypothesis that $\set{I}$ is exceptional.
Therefore, when $\alpha$ is not divisible by~$p$,
the only choice for $\alpha$ is $2w-1$, and hence $H = \langle 2w-1\rangle$. This completes the proof of the claim.

Let $\set{I}_1, \set{I}_2,\ldots, \set{I}_E$ be the exceptional codewords in~$\C$, and $H_j$ be the stabilizer of $d(\set{I}_j)$, for $j=1,2,\ldots, E$. It follows from the claim that
\[ \bigcup_{j=1}^E H_j \subseteq \langle 2w-1 \rangle \cup  \langle p \rangle.
\]

The same argument in the proof of Theorem~\ref{thm:extend} shows that at most $2w-2 - \varphi(2w-1)$ non-zero elements in $\langle p \rangle$, which is a subgroup in $\mathbb{Z}_{p(2w-1)}$ of size $2w-1$, belong to $H_j$ for some $j$.
Hence,
\begin{align}
 \sum_{j=1}^E (|H_j|-1 ) & = (p - 1) + (2w-2 - \varphi(2w-1)) \notag \\
& \leq p+ 2w - 3 -\varphi(2w-1). \label{eq:last}
\end{align}

Next, we note that $p$ and $2w-1$  are both relatively prime with $w$, hence $L =p (2w-1)$ is also relatively prime with $w$. Thus, as a divisor of $L$,  $|H_j|$ is relatively prime with $w$ for all $j=1,2,\ldots, E$. Since $|\set{I}_j + H_j|$ is an integral multiples of $|H_j|$, we have  $$|\set{I}_j + H_j| > w = |\set{I}_j|.$$
Recall that $\Delta_j$ in Theorem~\ref{lemma} is defined as $\Delta_j := |\set{I}_j + H_j| - w$.
We thus have $\Delta_j \geq 1$ for $j=1,2,\ldots, E$.

By Theorem~\ref{lemma}, we obtain the following upper bound on code size
\begin{align*}
|\C| & \leq \left\lfloor
\frac{L-1 + \sum_{j=1}^E (|H_j|-1 - 2\Delta_j)}{2w-2}
\right\rfloor \\
& \leq \left\lfloor\frac{L-1+p+2w-3-\varphi(2w-1)- 2E}{2w-2} \right\rfloor.
\end{align*}
Note that in the last equality, we have replaced $\sum_j (|H_j|-1)$ by~\eqref{eq:last} and  each $\Delta_j$ by~1.
After substituting $L$ by $p(2w-1)$, we obtain
\[
 |\C| \leq  p+ \left\lfloor \frac{2w+2p-4 -\varphi(2w-1)-2E}{2w-2} \right\rfloor.
\]
Since $2p \leq 2w + \varphi(2w-1)$ by assumption, we have
\[
|\C| \leq p + \left\lfloor \frac{4w-4-2E}{2w-2} \right\rfloor \leq p+1.
\]
In the last inequality, we have used the fact that $E\geq 1$.
This completes the proof of Theorem~\ref{thm:CRT}.
\end{proof}

Some new values of $M(L,w)$ determined by Theorem~\ref{thm:CRT} is shown in Table~\ref{table:CRT}.

\begin{table}
\[
\begin{array}{|c|c|c|} \hline
L & w & M(L,w) \\ \hline \hline
15 & 3  & 4 \\ \hline
35 &4  & 6 \\ \hline
45 &5  & 6 \\ \hline
63 &5  & 8 \\ \hline
77 &6  & 8 \\ \hline
91 &7  & 8 \\ \hline
165&8  & 12 \\ \hline
187&9  & 12 \\ \hline
221&9  & 14 \\ \hline
\end{array}
\]
\caption{The number of codewords of some optimal CACs from Theorem~\ref{thm:CRT}.}
\label{table:CRT}
\end{table}

\section{Conclusion}
We derive an upper bound for the size of CAC.  This is the first general bound which is applicable to any number of active users. For fixed Hamming weight $w$, the upper bound increases approximately with slope $(2w-2)^{-1}$ as a function of length~$L$. The upper bound is applied to some existing constructions of CAC, and many new values of $M(L,w)$  are determined.




\end{document}